\documentstyle[12pt]{article}

\newcommand{\ba}{\begin{eqnarray}}
\newcommand{\ea}{\end{eqnarray}}
\begin{document}
\pagestyle{plain}
\begin{center}
\huge
Algebraic-eikonal approach to medium energy proton scattering
from odd-mass nuclei
\vspace{1cm}\\
\Large
R. Bijker$^{1}$ and A. Frank$^{1,2}$
\normalsize
\vspace{1cm}\\
$^{1}$ Instituto de Ciencias Nucleares, U.N.A.M,\\
Apartado Postal 70-543, 04510 M\'{e}xico D.F., M\'{e}xico
\vspace{1cm}\\
$^{2}$ Instituto de F\'isica, Laboratorio de Cuernavaca,\\
Apartado Postal 139-B, 62191 Cuernavaca, Morelos, M\'exico
\vspace{1cm}
\vspace{1cm}\\
Keyword Abstract: Proton scattering.
Algebraic-eikonal approach. Odd-mass nuclei.
\end{center}

\begin{abstract}
We extend the algebraic-eikonal approach to medium
energy proton scattering from odd-mass nuclei by combining
the eikonal approximation for the scattering with a
description of odd-mass nuclei in terms of the interacting
boson-fermion model. We derive closed expressions for the
transition matrix elements for one of the dynamical symmetries
and discuss the interplay between collective and single-particle
degrees of freedom in an application to elastic and inelastic
proton scattering from $^{195}$Pt.
\end{abstract}

\section{Introduction}

High energy scattering of nucleons from a collective nucleus
involves the excitation of a large number of strongly coupled
states as well as that of virtual intermediate states.
The effects of channel coupling become particularly important for
large momentum transfer, where the scattering cannot be treated
in DWBA, but requires calculations to higher order in the channel
coupling \cite{AMS}. The standard approach is that of a coupled
channels calculation, which becomes complicated when the
number of channels that have to be included is large.
An alternative to this approach is provided
by the eikonal approximation, which in combination with the
Interacting Boson Model (IBM) \cite{IBM} allows the calculation of
the scattering to all orders in closed form for even-even nuclei
\cite{AMS,GOAS,GW}. This method, also known as the algebraic-eikonal
approach, combines the strengths of an algebraic description of
the target dynamics with those of the eikonal approximation
in the adiabatic limit. In the case of odd-mass nuclei
the complications of the channel coupling approach are even greater,
because of the interplay between collective and single-particle
degrees of freedom and the increase in the number of open channels
in the system. On the other hand, the Interacting Boson-Fermion Model
(IBFM) has provided a
tractable model of odd-mass nuclei and its usefulness for the
classification and understanding of nuclear structure data has
been tested in numerous ways \cite{IBFM}. Among the most interesting
aspects of this model is the possibility of the occurrence of
boson-fermion dynamical symmetries as well as supersymmetries,
the latter of which link the properties of neighboring nuclei.

The purpose of this brief report is twofold. First
we show that the algebraic-eikonal approach for medium
energy proton scattering can be extended to odd-mass nuclei in a
simple fashion by describing the target nucleus in terms of the IBFM.
Next we consider the particular case of one of the
dynamical symmetries of the IBFM, the $SO(6) \otimes SU(2)$ limit,
to derive closed expressions for
the transition matrix elements. As an example we discuss
the application to elastic and inelastic proton scattering
from $^{195}$Pt, which is considered to be a paradigm of
dynamical boson-fermion symmetry, both in terms of energy systematics,
electromagnetic decay properties and single-particle transfer
\cite{BI}. We consider in particular
the interplay between the coupling to collective
and single-particle degrees of freedom for the excitation of
the lowest negative parity states in $^{195}$Pt by medium
energy protons.

\section{Eikonal Approximation}

For medium and high energy proton-nucleus scattering the eikonal
approach is a good approximation for elastic and inelastic scattering.
The hamiltonian is in general given by
\ba
H=\frac{\hbar^2 k^2}{2m} + H_{t}(\xi) + V(\vec{r},\xi) ~,
\ea
where $H_{t}$ describes the dynamics of the target nucleus and
$V(\vec{r},\xi)$ represents the proton-nucleus interaction.
The projectile coordinate $\vec{r}$ is measured from the center of
mass of the target. The internal coordinates of the target nucleus are
collectively denoted by $\xi$. If the kinetic energy of the projectile
is much larger than the interaction strength, and is also
sufficiently large that the projectile wavelength is small compared
with the range of variation of the potential, one may use the eikonal
approximation to describe the scattering. If, in addition, the
projectile energy is large compared with the nuclear excitation
energies, one can neglect $H_{t}$ (adiabatic limit). Under these
approximations the scattering amplitude for scattering a projectile
with initial momentum $\vec{k}$ from an initial state
$|i \rangle$ to final momentum $\vec{k}^{\prime}$ and a final state
$|f \rangle$ is given by
\ba
A_{fi}(\vec{q}) &=&
\frac{k}{2 \pi i} \int \mbox{d}^2 b \; e^{i\vec{q}\cdot\vec{b}} \,
\langle f|e^{i \chi(\vec{b},\xi)}-1|i \rangle ~,
\ea
where $\vec{q}=\vec{k}^{\prime}-\vec{k}$ is the momentum transfer and
$\chi(\vec{b},\xi)$ is the eikonal phase that
the projectile acquires as it goes by the target
\ba
\chi(\vec{b},\xi) &=& -\frac{m}{\hbar^{2}k} \int^{\infty}_{-\infty}
\mbox{d} z \; V(\vec{r},\xi) ~. \label{chi}
\ea
In the derivation of the scattering amplitude the projectile coordinate
is written as $\vec{r}=\vec{b}+\vec{z}$, where the impact parameter
$\vec{b}$ is perpendicular to the $z$-axis, which is chosen along
$\hat z=(\vec{k}+\vec{k}^{\prime})/|\vec{k}+\vec{k}^{\prime}|$.
In the eikonal approximation the scattering amplitude is
expressed in terms of an integral over a two-dimensional impact
parameter rather than as a sum over partial waves.

For medium energy proton-nucleus scattering from even-even nuclei
the eikonal approximation has been applied successfully to elastic
and inelastic scattering to forward angles \cite{AMS,GOAS,GW}.
This procedure can be extended to odd-mass nuclei by considering
the coupling of the projectile to both the collective and
the single-particle degrees of freedom.
If the range of the projectile-nucleus interaction is short
compared to the size of the nucleus, the potential can be expressed
in terms of the projectile-nucleon forward scattering amplitude $f$
\cite{AMS}
\ba
V(\vec{r},\xi) &=& - \frac{2 \pi \hbar^2 f}{m} \, \Bigl[
\rho(r)+ \left[ Q_B(r,\xi) + Q_F(r,\xi) \right]
\cdot Y_{2}(\hat{r}) \Bigr] ~.
\ea
Here $\rho(r)$ is the nuclear density for the distorting
or optical potential, and $Q_{B}(r,\xi)$ and $Q_{F}(r,\xi)$ denote
contributions from the quadrupole coupling of the projectile to
the collective (bosonic) and single-particle (fermionic)
degrees of freedom of the odd-mass nucleus.

For a strongly absorbing probe the scattering is dominated by
peripheral collisions, which allows one to keep only the leading
order term in the expansion of the spherical harmonic around
$\theta=\pi/2$,
\ba
Y_{2\mu}(\hat{r}) &=& Y_{2\mu}(\hat{b}) + {\cal O}(z/r) ~.
\ea

The calculation of the nuclear matrix elements to all orders in
$\chi$ is a complicated task. The use of algebraic models to
describe the nuclear excitations makes such a calculation feasible.

\section{The Interacting Boson-Fermion Model}

In the IBFM the collective and single-particle quadrupole
operators are given by
\ba
Q_{B,\mu}(r,\xi) &=&
  \alpha_1(r) \, [s^{\dagger} \tilde{d} + d^{\dagger} s]^{(2)}_{\mu}
+ \alpha_2(r) \, [d^{\dagger} \tilde{d}]^{(2)}_{\mu} ~,
\nonumber\\
Q_{F,\mu}(r,\xi) &=&
\sum_{j \leq j^{\prime}} \alpha_{j j^{\prime}}(r) \,
[a^{\dagger}_j \tilde{a}_{j^{\prime}} + (-1)^{j-j^{\prime}} \,
 a^{\dagger}_{j^{\prime}} \tilde{a}_j]^{(2)}_{\mu}
/ (1+\delta_{jj^{\prime}}) ~,
\ea
with $\tilde{d}_{\mu}=(-1)^{2-\mu} d_{-\mu}$ and
$\tilde{a}_{j,\mu}=(-1)^{j-\mu} a_{j,-\mu}$.
Since the quadrupole operators are linear in the generators
of the symmetry group of the IBFM, $G=U_B(6) \otimes U_F(m)$ (with
$m=\sum_j (2j+1)$), the eikonal transition matrix elements can be
interpreted as group elements of $G$. They are thus a generalization
of the Wigner ${\cal D}$-matrices for the rotation group.
In general, these representation matrix elements can be calculated
exactly (albeit numerically) to all orders in the projectile-nucleus
coupling strength, either with or without the
peripheral approximation. This holds for any collective nucleus,
either spherical, deformed, $\gamma$ unstable or an
intermediate situation between them.
The general result can be expressed in terms of a five-dimensional
integral for the collective part \cite{Wenes} and a contribution
from the single-particle part, which is easily obtained for a single
nucleon. In the peripheral approximation the expression for the
scattering amplitude for scattering from an initial state
$|i \rangle=|\alpha,J,M \rangle$ to a final state
$|f \rangle=|\alpha^{\prime},J^{\prime},M^{\prime} \rangle$
reduces to a one-dimensional integral over the impact parameter
\ba
A_{fi}(\vec{q}) &=&
\frac{k}{i} \; i^{M-M^{\prime}} \int_{0}^{\infty} b \mbox{d} b \;
J_{M-M^{\prime}}(qb) \, \left[ e^{i \chi_{opt}(b)}
\sum_{M^{\prime \prime}}
{\cal D}^{(J^{\prime})}_{M^{\prime}M^{\prime \prime}}(\hat{q}) \right.
\nonumber\\
&& \times \left. \langle \alpha^{\prime},J^{\prime},M^{\prime \prime} |
e^{i [\chi_B(b,\xi) + \chi_F(b,\xi)]}
| \alpha,J,M^{\prime \prime} \rangle \,
{\cal D}^{(J)}_{M^{\prime \prime}M}(-\hat q)
- \delta_{fi} \right] ~. \label{afi}
\ea
The projectile distorted wave is given by
\ba
\chi_{opt}(b) &=& \frac{2 \pi f}{k} \int^{\infty}_{-\infty}
\mbox{d} z \; \rho(r) ~,
\ea
and the boson and fermion eikonal phases by
\ba
\chi_B(b,\xi) &=&
  g_1(b) \, [s^{\dagger} \tilde{d} + d^{\dagger} s]^{(2)}_0
+ g_2(b) \, [d^{\dagger} \tilde{d}]^{(2)}_0 ~,
\nonumber\\
\chi_F(b,\xi) &=& \sum_{j \leq j^{\prime}} g_{j j^{\prime}}(b) \,
[a^{\dagger}_j \tilde{a}_{j^{\prime}} + (-1)^{j-j^{\prime}}
 a^{\dagger}_{j^{\prime}} \tilde{a}_j]^{(2)}_0
/ (1+\delta_{jj^{\prime}}) ~,
\ea
with the eikonal profile functions
\ba
g_1(b) &=& \frac{2 \pi f}{k} \int^{\infty}_{-\infty}
\mbox{d} z \; \alpha_1(r) \sqrt{5/4\pi} ~,
\nonumber\\
g_2(b) &=& \frac{2 \pi f}{k} \int^{\infty}_{-\infty}
\mbox{d} z \; \alpha_2(r) \sqrt{5/4\pi} ~,
\nonumber\\
g_{j j^{\prime}}(b) &=& \frac{2 \pi f}{k} \int^{\infty}_{-\infty}
\mbox{d} z \; \alpha_{j j^{\prime}}(r) \sqrt{5/4\pi} ~.
\ea
Hence the only representation matrix elements that are needed
are those that depend on the $z$-component of the quadrupole operator.
Without the peripheral approximation the other components have to be
included as well. In the special case of a dynamical symmetry
the matrix elements appearing in Eq.~(\ref{afi}) can be derived in
closed form. In \cite{GOAS} the results for each of the dynamical
symmetries of the IBM for even-even nuclei were analyzed.
Here we present the first such study for odd-mass nuclei.

One of the best examples of dynamical symmetries in odd-even nuclei
is provided by the low lying negative parity states of $^{195}$Pt,
which have been analyzed successfully in terms of the
$U(6) \otimes SU(2) \supset SO(6) \otimes SU(2)$ limit of the IBFM
\cite{BI}. The odd neutron in $^{195}$Pt occupies the $3p_{1/2}$,
$3p_{3/2}$ and $2f_{5/2}$ shell model orbits with $n=5$, which are
treated in a pseudo-spin coupling scheme as the $3\bar{s}_{1/2}$,
$2\bar{d}_{3/2}$ and $2\bar{d}_{5/2}$ pseudo-orbits with
$\bar{n}=n-1=4$. In this limit the quadrupole operators are
\ba
\hat Q_{B,\mu} &=& [s^{\dagger} \tilde{d} + d^{\dagger} s]^{(2)}_{\mu} ~,
\nonumber\\
\hat Q_{F,\mu} &=&
-\sqrt{4/5} \, [a^{\dagger}_{1/2} \tilde{a}_{3/2}
- a^{\dagger}_{1/2} \tilde{a}_{3/2}]^{(2)}_{\mu}
-\sqrt{6/5} \, [a^{\dagger}_{1/2} \tilde{a}_{5/2}
+ a^{\dagger}_{1/2} \tilde{a}_{5/2}]^{(2)}_{\mu} ~. \label{qbf}
\ea
Therefore there are only two independent eikonal profile functions,
one for the collective part
\ba
g_1(b) &=& \epsilon_B(b) ~,
\nonumber\\
g_2(b) &=& 0 ~,
\ea
and one for the single-particle part
\ba
g_{1/2,3/2} &=& -\sqrt{4/5} \, \epsilon_F(b) ~,
\nonumber\\
g_{1/2,5/2} &=& -\sqrt{6/5} \, \epsilon_F(b) ~.
\ea
The latter values are a consequence of the pseudo-spin coupling
scheme for the single-particle orbits \cite{BI}.

The classification scheme and the structure of the wave functions
are discussed in detail in \cite{BI}. The eigenstates are labeled by
$|[N_1,N_2],(\sigma_1,\sigma_2,\sigma_3),(\tau_1,\tau_2),L,J^P \rangle$.
Here $L$ denotes the pseudo-orbital angular momentum, which is a
combination of the core angular momentum, $R$, and the
pseudo-orbital part of the single-particle angular momenta, $l$.
In this classification scheme the states occur in pseudo-spin
doublets with total angular momentum $J=L \pm 1/2$ for $L > 0$ or in
singlets with $J=1/2$ for $L=0$.
Some low lying excitations are listed in Table~\ref{BE2} together
with their $B(E2)$ values to the ground state,
$|[N+1,0],(N+1,0,0),(0,0),0,1/2^-\rangle$~.

The eikonal transition matrix elements
can be obtained by expanding the wave functions of the initial
and final states into the direct product of collective (boson) and
single-particle (fermion) basis states \cite{BI}. The boson part
of the transition matrix element can be expressed in terms of
Gegenbauer polynomials using similar techniques as for proton
scattering from even-even nuclei in the $SO(6)$ limit \cite{GOAS}.
The fermion part is easy to evaluate for a single uncoupled nucleon.
For elastic transitions we find
\ba
U_{\mbox{el}}(b) &=& \frac{4!N!}{(N+2)(N+4)!}
\left[ (N+2) \cos \epsilon_{F} \, C^{(3)}_{N}(\cos \epsilon_{B}) \right.
\nonumber\\
&&\hspace{1cm} - \left.
\left[ (N+4) \cos (\epsilon_{B}-\epsilon_{F}) -2 \right] \,
C^{(3)}_{N-1}(\cos \epsilon_{B}) \right] ~.
\ea
The matrix elements for transitions to excited states
can be derived in a similar way.
The matrix element to the first excited pseudo-spin doublet
in the ground state band $[N+1,0],(N+1,0,0),(1,0),2$ is given by
\ba
U_{2_1}(\epsilon_{B},\epsilon_{F}) &=&
\frac{4!N!}{(N+2)(N+4)!} \sqrt{\frac{1}{5(N+1)(N+5)}}
\nonumber\\
&& \times \left[ (3N+2)(N+5) \, i \sin \epsilon_{F} \,
C^{(3)}_{N}(\cos \epsilon_{B}) \right.
\nonumber\\
&&+ 3(N+4)(N+5) \, i \sin (\epsilon_{B}-\epsilon_{F}) \,
C^{(3)}_{N-1}(\cos \epsilon_{B})
\nonumber\\
&&+ 12(N-1) \, i \sin \epsilon_{B} \cos \epsilon_{F} \,
C^{(4)}_{N-1}(\cos \epsilon_{B})
\nonumber\\
&&+ \left. 12 i \sin \epsilon_{B}
\left[ 5-(N+4) \cos (\epsilon_{B}-\epsilon_{F}) \right] \,
C^{(4)}_{N-2}(\cos \epsilon_{B}) \right] ~.
\ea
The first excited state in $^{195}$Pt belongs to the pseudo-spin
doublet characterized by $[N,1],(N,1,0),(1,0),2$. The transition
matrix element to this doublet is given by
\ba
U_{2_2}(\epsilon_{B},\epsilon_{F}) &=&
\frac{4!N!}{(N+4)!} \sqrt{\frac{2(N-1)!}{5(N+3)!}}
\left[ N(N-1) \, i \sin \epsilon_{F} \,
C^{(3)}_{N}(\cos \epsilon_{B}) \right.
\nonumber\\
&&+ (N+4)(N-5) \, i \sin (\epsilon_{B}-\epsilon_{F}) \,
C^{(3)}_{N-1}(\cos \epsilon_{B})
\nonumber\\
&&- 6(N-1) \, i \sin \epsilon_{B} \cos \epsilon_{F} \,
C^{(4)}_{N-1}(\cos \epsilon_{B})
\nonumber\\
&&- \left. 6 i \sin \epsilon_{B}
\left[ 5-(N+4) \cos (\epsilon_{B}-\epsilon_{F}) \right] \,
C^{(4)}_{N-2}(\cos \epsilon_{B}) \right] ~.
\ea
In the peripheral approximation the
states with odd values of the (pseudo-orbital) angular momentum $L$
(such as the pseudo-spin doublet with $L=1$ in Table~\ref{BE2})
cannot be excited
\ba
U_{1}(b) &=& 0 ~.
\ea

For $\epsilon_{B}=\epsilon_{F}=\epsilon$ the quadrupole operator
becomes a generator of the $SO(6)$ group and can therefore connect
only states belonging to the same $SO(6)$ representation. Using
a recurrence relation for the Gegenbauer polynomials
\ba
n C^{(p)}_{n}(x) &=& 2p \left[
x C^{(p+1)}_{n-1}(x) - C^{(p+1)}_{n-2}(x) \right] ~,
\ea
it is easy to show that in this case the above transition
matrix reduce to the expressions derived for proton scattering
from even-even nuclei with $SO(6)$ symmetry \cite{GOAS}
\ba
U_{\mbox{el}}(b) &=&
\frac{3!(N+1)!}{(N+4)!} \, C^{(2)}_{N+1}(\cos \epsilon) ~,
\nonumber\\
U_{2_1}(b) &=&
\frac{4!N!}{(N+4)!} \sqrt{\frac{5(N+1)}{N+5}} \,
(i \sin \epsilon) \, C^{(3)}_{N}(\cos \epsilon) ~,
\nonumber\\
U_{2_2}(b) &=& 0 ~.
\ea

\section{Application to $^{195}$Pt}

The mass region of the Pt isotopes is known as a complex
transitional region of the nuclear mass table between deformed
and gamma-unstable nuclei. Nevertheless, some of the best examples of
dynamical symmetries of the IBM in even-even nuclei ($^{194,196}$Pt)
and of the IBFM in odd-mass nuclei ($^{195}$Pt) are found in this
mass region. In \cite{evenpt} the even-even Pt nuclei were studied
in proton scattering. Here we present the first results of
calculations for the scattering of 800 MeV protons from $^{195}$Pt.

The low-lying negative parity states of $^{195}$Pt show a very
small splitting between states belonging to the same pseudo-spin
doublet (too small to be resolved experimentally in proton
scattering) (see Table~1).
Therefore we calculate the differential cross section (d.c.s.)
for a given pseudo-spin doublet which is summed over the contributions
of the individual states. Under the assumption of a pseudo-spin
coupling scheme for the single-particle orbits, the angular
distributions for the excitation of the individual states of
a pseudo-spin multiplet characterized by $L$, are identical up to a
statistical factor
\ba
\frac{\mbox{d} \sigma(0,1/2 \rightarrow L,J|q)}{\mbox{d}\Omega}
&=& \frac{2J+1}{2(2L+1)}
\frac{\mbox{d} \sigma(0 \rightarrow L|q)}{\mbox{d}\Omega} ~.
\ea
Hence the summed d.c.s. shows the same dependence on momentum transfer
as the indivial contributions. However, if the pseudo-spin assumption
would be broken significantly, the predicted narrow oscillations are
likely to be washed out.

In Figure~\ref{Pt195} we show the d.c.s. for the scattering of 800 MeV
protons from $^{195}$Pt which, as mentioned before, is the best known
example of an odd-mass nucleus with $SO(6) \otimes SU(2)$ symmetry.
The three curves represent elastic scattering and the
excitation of two low-lying pseudo-spin doublets
with $L=2$ (see Table~\ref{BE2}). In the calculations we assume
a Woods-Saxon form for the nuclear density
with a nuclear radius of $1.2 A^{1/3}=6.96$ fm and a diffusivity of
0.75 fm, normalized to the total number of nucleons $A=195$. For the
collective transition density we use the derivative of the nuclear
density (Tassie form) and for the single-particle transition density a
product of radial wave functions for the pseudo-oscillator orbits.
We note that
in the pseudo-spin coupling scheme there is a single transition
density for the fermion quadrupole operator of Eq.~(\ref{qbf}).
The transition densities are normalized to the $B(E2)$ values in
Table~\ref{BE2}. The forward proton-nucleon scattering amplitude is
$f=ik\sigma/4\pi$, in which the isospin averaged proton-nucleon
cross section was taken as $\sigma=46(1+0.38i)$ mb \cite{GOAS}.

The d.c.s. of Figure~\ref{Pt195} incorporate the effects of the
interplay of the coupling to the collective and the single-particle
degrees of freedom in the target nucleus.
In Figures~\ref{Pt0}--\ref{Pt2}
we gauge these effects by comparing the results
of the full calculation (solid lines) with those of a calculation
in which the coupling to the single-particle degrees of freedom
is turned off (dashed lines).
The elastic scattering ($L=0$) is completely determined by the
collective part. Whereas the excitation of the $L=2$ doublet
which belongs to the ground band $[7,0],(7,0,0)$ is still
largely determined by the collective part, for the excitation of
the $L=2$ doublet with $[6,1],(6,1,0)$ we predict a large contribution
from the single-particle part as well. Note that the scale
is logarithmic, so there is almost a factor of two difference between
the two curves in Figure~\ref{Pt2}.

\section{Summary and conclusions}

In this brief report we have presented an extension of the
algebraic-eikonal approach to medium and high energy proton scattering
from odd-mass nuclei described by the Interacting Boson-Fermion Model.
The algebraic structure of the IBFM makes it possible to calculate
the eikonal transition matrix elements exactly to all orders in the
projectile-nucleus coupling strength. This holds for any type of
collective nucleus, either spherical, deformed, $\gamma$ unstable
or any intermediate situation between them.

In the special case of a dynamical symmetry all transition matrix
elements of interest can be obtained in closed
form. We have discussed in particular an application to the
negative parity states in $^{195}$Pt, which are described
in terms of the $SO(6) \otimes SU(2)$ limit of the IBFM. The analytic
expressions for the eikonal transition matrix elements allow the
study of various effects in a straightforward way.
As an example, we showed that whereas the excitation of the states
belonging to the $[N+1,0],(N+1,0,0)$
ground band are largely dominated by coupling to the collective
degrees of freedom, the excitation of the $[N,1],(N,1,0)$ side band
depends sensitively on the interplay between the coupling to the
single-particle and collective degrees of freedom.
It would be of interest to experimentally test the
pseudo-spin symmetry assumption discussed
in this report through proton scattering from $^{195}$Pt.

Finally, we remark that the formalism presented here can be used to
derive closed expressions for the transition matrix elements for
other dynamical symmetries of the IBFM, and can be extended to
odd-odd nuclei as well.

\section*{Acknowledgements}

This work was supported in part by CONACyT, M\'exico under project
400340-5-3401E, DGAPA-UNAM under project IN105194 and the European
Community under contract nr CI1$^{\ast}$-CT94-0072.

\clearpage

\begin{table}
\centering
\caption[]{$B(E2)$ values leading to the ground state in $^{195}$Pt,
calculated with $e_B=0.184$ (eb) and $e_F=-0.257$ (eb).
The number of bosons is $N=6$.}
\vspace{30pt}
\label{BE2}
\begin{tabular}{cclcc}
\hline
& & & & \\
Initial state & $B(E2)$ & $J^P$(keV) &
\multicolumn{2}{c}{$B(E2)$(e$^2$b$^2$)} \\
& th & & exp \cite{Pt} & calc \\
& & & & \\
\hline
& & & & \\
\, [7,0],(7,0,0),(1,0),2,$J^P$ &
$(Ne_B+e_F)^2 \frac{N+5}{5(N+1)}$ &
$3/2^-$(211) & 0.240(25) & 0.225 \\
& & $5/2^-$(239) & 0.210(23) & 0.225 \\
& & & & \\
\, [6,1],(6,1,0),(1,0),2,$J^P$ &
$(e_B-e_F)^2 \frac{2N(N+3)}{5(N+1)(N+2)}$ &
$3/2^-$(99) & 0.085(20) & 0.075 \\
& & $5/2^-$(130) & 0.066(10) & 0.075 \\
& & & & \\
\, [6,1],(6,1,0),(1,1),1,$J^P$ & 0 & $1/2^-$ & & 0 \\
& & $3/2^-$(199) & 0.019(5) & 0 \\
& & & & \\
\hline
\end{tabular}
\end{table}

\clearpage

\begin{figure}
\caption[]{Differential cross sections in mb/sr for elastic
scattering (solid line) and the excitation
of the pseudo-orbital doublets with $[7,0],(7,0,0),(1,0),L=2,J$
(dashed line) and $[6,1],(6,1,0),(1,0),L=2,J$ (dotted line)
in $^{195}$Pt by 800 MeV protons.}
\label{Pt195}
\end{figure}

\begin{figure}
\caption[]{Elastic differential cross section in mb/sr calculated
with and without the coupling to the single-particle degrees
of freedom (solid and dashed lines, respectively).}
\label{Pt0}
\end{figure}

\begin{figure}
\caption[]{Differential cross section in mb/sr for the excitation
of the pseudo-orbital doublet with $[7,0],(7,0,0),(1,0),L=2,J$
calculated with and without the coupling to the single-particle
degrees of freedom (solid and dashed lines, respectively).}
\label{Pt1}
\end{figure}

\begin{figure}
\caption[]{Differential cross section in mb/sr for the excitation
of the pseudo-orbital doublet with $[6,1],(6,1,0),(1,0),L=2,J$
calculated with and without the coupling to the single-particle
degrees of freedom (solid and dashed lines, respectively).}
\label{Pt2}
\end{figure}

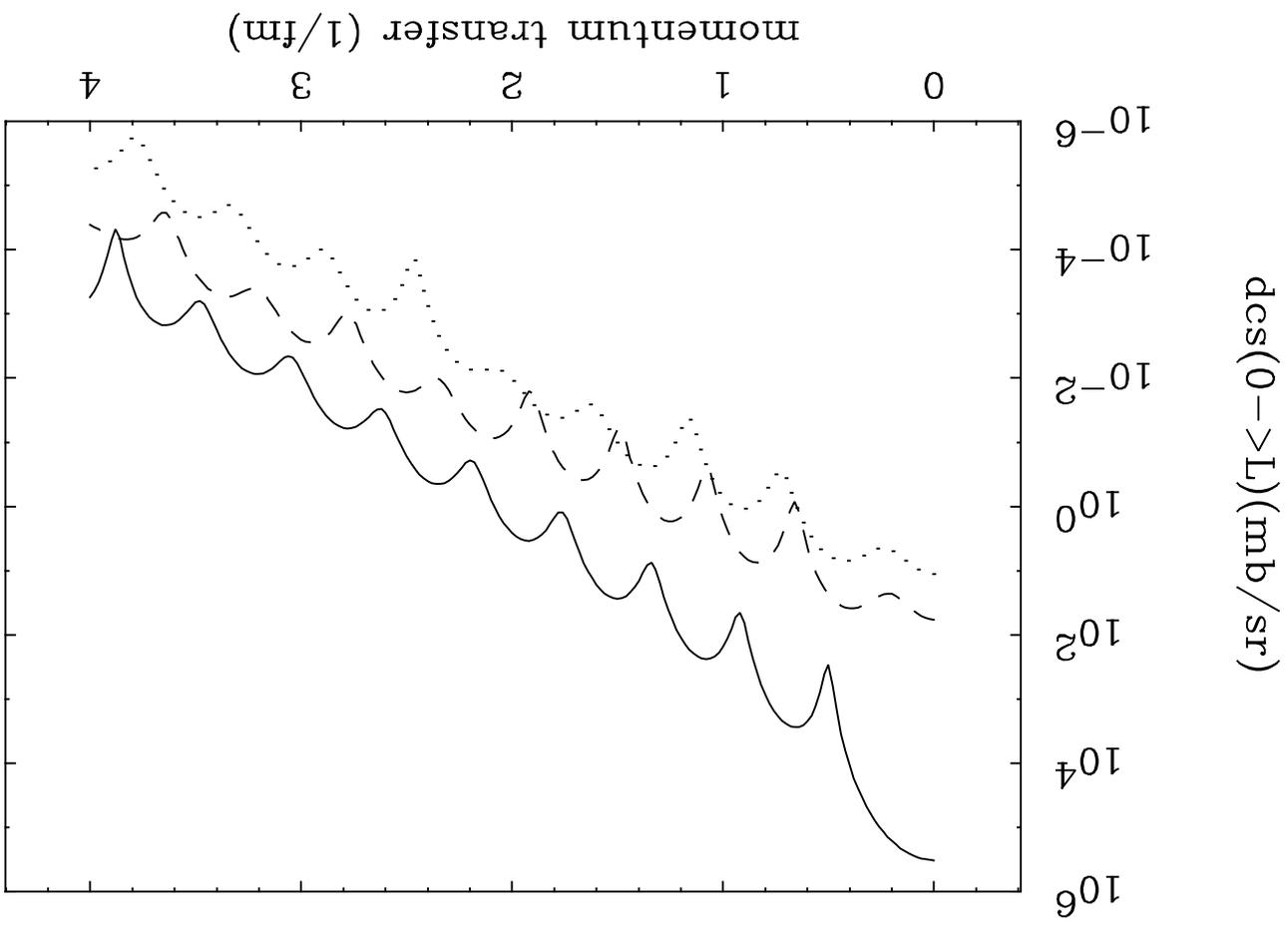
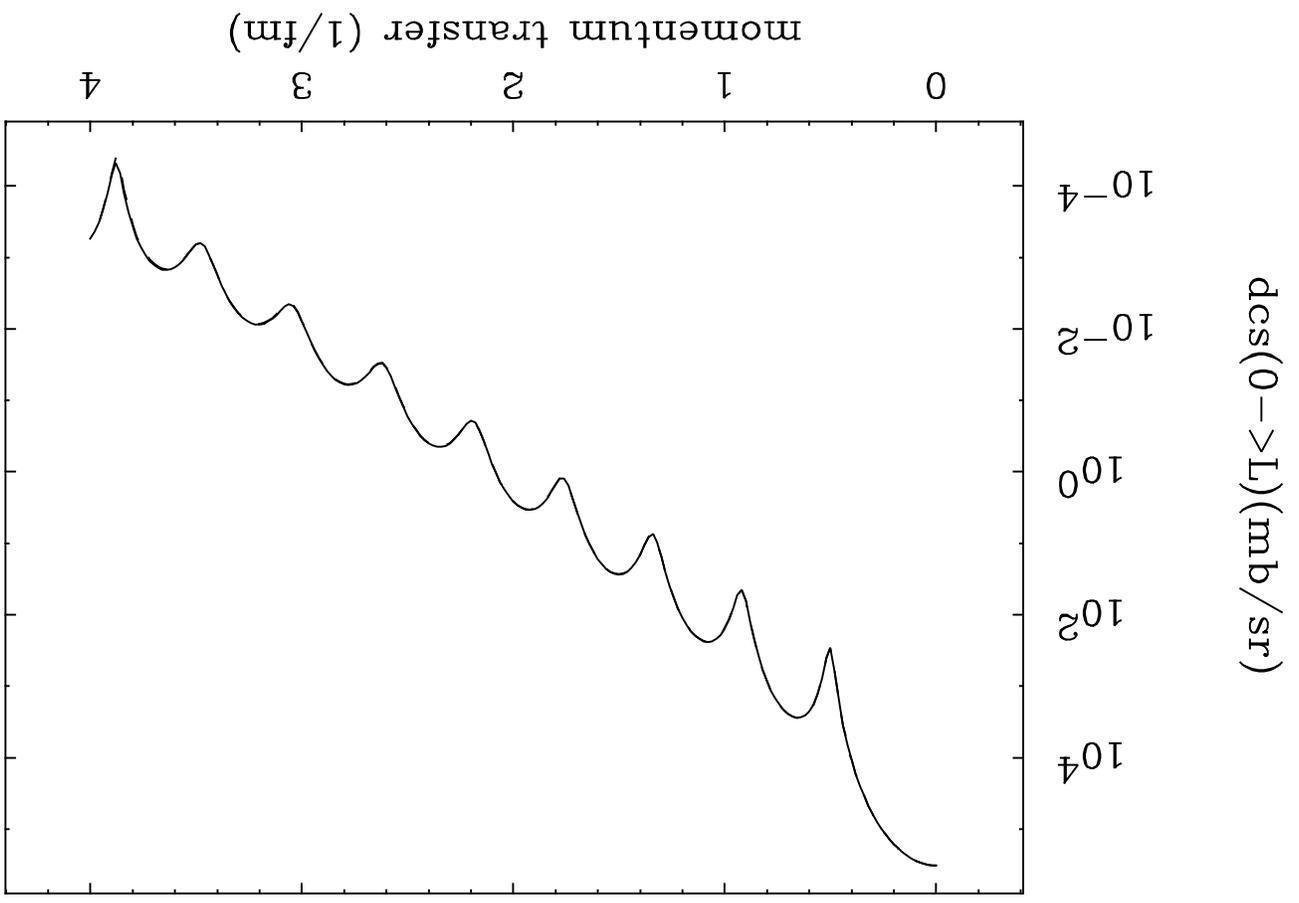
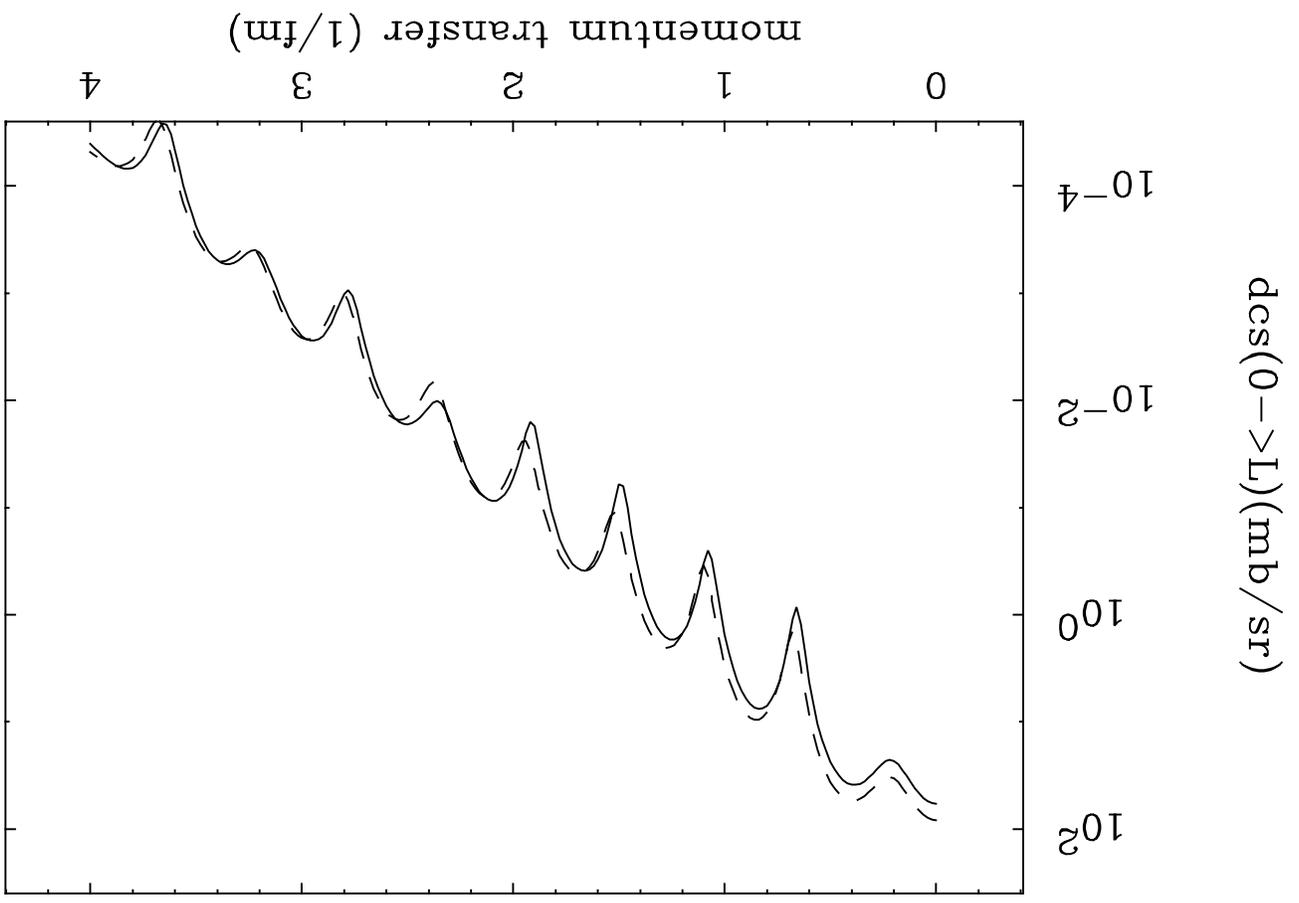
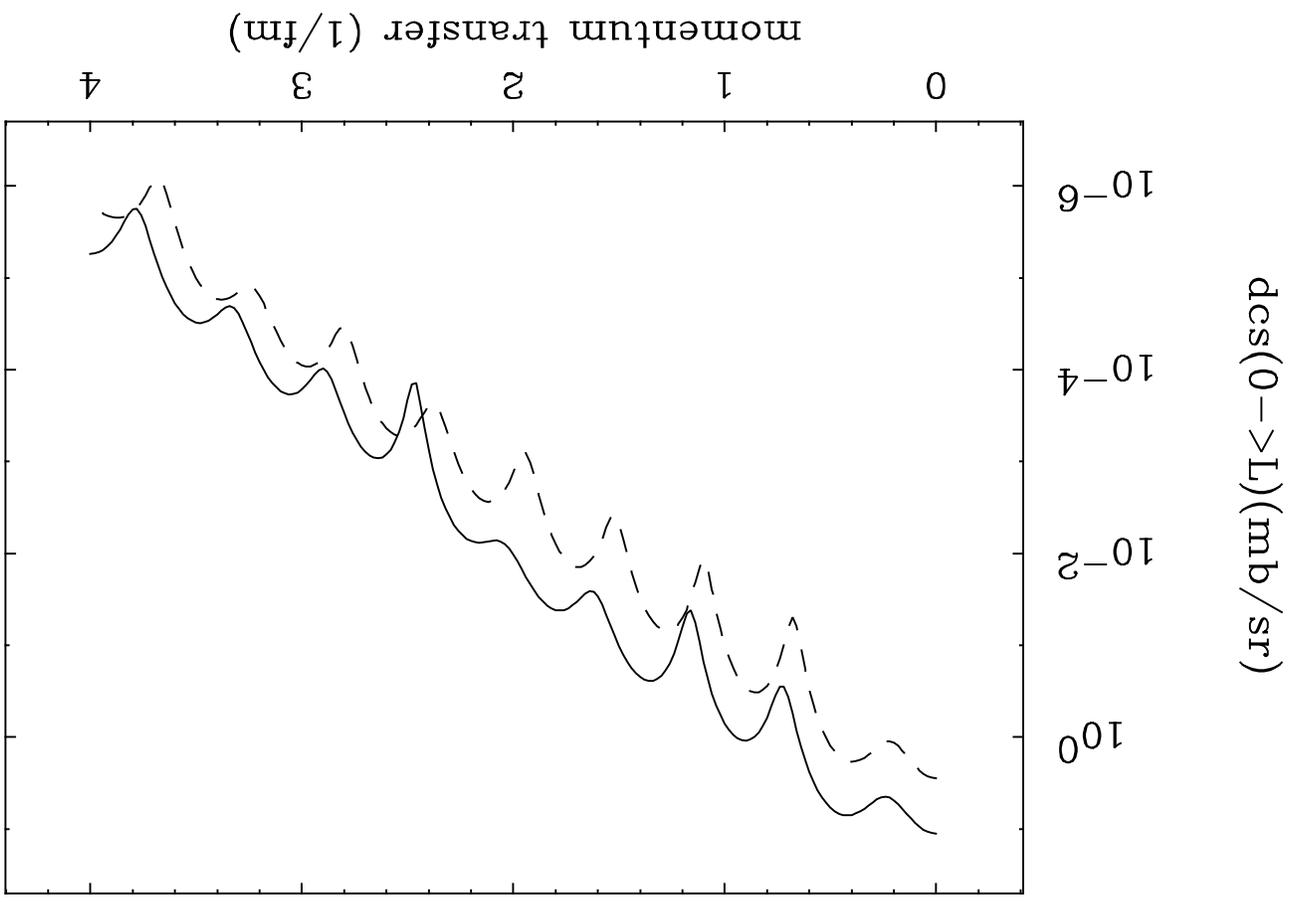
\end{document}